\documentclass[11pt,a4paper]{article}

\usepackage[utf8x]{inputenc}
\usepackage[T1]{fontenc}
\usepackage{color}
\usepackage{amssymb}
\usepackage{amsmath}
\usepackage{graphicx}
\usepackage{mathtools}
\usepackage{ragged2e}
\usepackage{psfrag}

\usepackage{dcolumn}% Align table columns on decimal point
\usepackage{bm}
\usepackage{graphicx,empheq}
\usepackage{verbatim} 
\usepackage[english]{babel}
\usepackage{hyperref}

\hypersetup{
citecolor=red,
colorlinks=true,
filecolor=red,
linkcolor=blue,
linktocpage=true,
urlcolor=blue
} 
\usepackage[titletoc,toc,title]{appendix}
\numberwithin{equation}{section}
\usepackage{cleveref}
\usepackage{cite}
\usepackage{epsfig}
\usepackage{float}
\usepackage{amsfonts}
\usepackage{enumitem}
\usepackage[font={footnotesize,it}]{caption}

\newcommand\trick[1]{}
\newcommand{\be}{\begin{equation}} 
\newcommand{\ee}{\end{equation}}
\newcommand{\eq}[1]{(\ref{#1})}
\newcommand{\bit}{\begin{itemize}}  \newcommand{\eit}{\end{itemize}}
\newcommand{\ben}{\begin{enumerate}}  \newcommand{\een}{\end{enumerate}}

\newcommand{\rf}[1]{(\ref{#1})}

\def\bd{\begin{document}}
\def\ed{\end{document}}
\def\bea{\begin{eqnarray}}
\def\eea{\end{eqnarray}}
\let\bm=\bibitem

\def\la{\langle}
\def\ra{\rangle}

\def\npb#1#2#3{Nucl. Phys. {\bf{B#1}} #3 (#2)}
\def\plb#1#2#3{Phys. Lett. {\bf{#1B}} #3 (#2)}
\def\prl#1#2#3{Phys. Rev. Lett. {\bf{#1}} #3 (#2)}
\def\prd#1#2#3{Phys. Rev. {D bf{#1}} #3 (#2)}
\def\cmp#1#2#3{Comm. Math. Phys. {\bf{#1}} #3 (#2)}
\def\cqg#1#2#3{Class. Quantum Grav. {\bf{#1}} #3 (#2)}
\def\nppsa#1#2#3{Nucl. Phys. B (Proc. Suppl.) {\bf{#1A}}#3 (#2)}
\def\ap#1#2#3{Ann. of Phys. {\bf{#1}} #3 (#2)}
\def\ijmp#1#2#3{Int. J. Mod. Phys. {\bf{A#1}} #3 (#2)}
\def\rmp#1#2#3{Rev. Mod. Phys. {\bf{#1}} #3 (#2)}
\def\mpla#1#2#3{Mod. Phys. Lett. {\bf A#1} #3 (#2)}
\def\jhep#1#2#3{J. High Energy Phys. {\bf #1} #3 (#2)}
\def\atmp#1#2#3{Adv. Theor. Math. Phys. {\bf #1} #3 (#2)}

\def\sst{\scriptscriptstyle}
\def\thetabar{\bar\theta}
\def\Tr{{\rm Tr}}
\def\one{\mbox{1 \kern-.59em {\rm l}}}

\def\a{\alpha}      \def\da{{\dot\alpha}}  \def\dA{{\dot A}}
\def\b{\beta}       \def\db{{\dot\beta}}
\def\g{\gamma}  \def\G{\Gamma}  \def\dc{{\dot\gamma}}
\def\d{\delta}  \def\D{\Delta}  \def\ddt{\dot\delta}
\def\e{\epsilon}
\def\ve{\varepsilon}
\def\uve{\upvarepsilon}
\def\f{\phi}    \def\F{\Phi}    \def\vvf{\f}
\def\vphi{\varphi}
\def\h{\eta}
\def\k{\kappa}
\def\l{\lambda} \def\L{\Lambda}
\def\m{\mu} \def\n{\nu}
\def\o{\omega}
\def\p{\pi} \def\P{\Pi}
\def\r{\rho}
\def\s{\sigma}  \def\S{\Sigma}
\def\t{\tau}
\def\th{\theta} \def\Th{\Theta} \def\vth{\vartheta}
\def\X{\Xeta}
\def\z{\zeta}

\def\na{\nabla}
\def\cA{{\cal A}} \def\cB{{\cal B}} \def\cC{{\cal C}}
\def\cD{{\cal D}} \def\cE{{\cal E}} \def\cF{{\cal F}}
\def\cG{{\cal G}} \def\cH{{\cal H}} \def\cI{{\cal I}}
\def\cJ{{\mathscr J}} \def\cK{{\cal K}} \def\cL{{\cal L}}
\def\cM{{\cal M}} \def\cN{{\cal N}} \def\cO{{\cal O}}
\def\cP{{\cal P}} \def\cQ{{\cal Q}} \def\cR{{\cal R}}
\def\cS{{\cal S}} \def\cT{{\cal T}} \def\cU{{\cal U}}
\def\cV{{\cal V}} \def\cW{{\cal W}} \def\cX{{\cal X}}
\def\cY{{\cal Y}} \def\cZ{{\cal Z}}
\def\ct{{\cal t}}

\def\ua{\underline{\alpha}}
\def\uc{\underline{\phantom{\alpha}}\!\!\!\gamma}
\def\um{\underline{\mu}}
\def\ud{\underline\delta}
\def\ue{\underline\epsilon}
\def\una{\underline a}\def\unA{\underline A}
\def\unb{\underline b}\def\unB{\underline B}
\def\unc{\underline c}\def\unC{\underline C}
\def\und{\underline d}\def\unD{\underline D}
\def\une{\underline e}\def\unE{\underline E}
\def\unf{\underline{\phantom{e}}\!\!\!\! f}\def\unF{\underline F}
\def\unm{\underline m}\def\unM{{\underline M}}
\def\unn{\underline n}\def\unN{{\underline N}}
\def\unp{\underline{\phantom{a}}\!\!\! p}\def\unP{\underline P}
\def\unq{\underline{\phantom{a}}\!\!\! q}
\def\unQ{\underline{\phantom{A}}\!\!\!\! Q}
\def\unH{\underline{H}}

\def\As {{A \hspace{-6.4pt} \slash}\;}
\def\bs {{b \hspace{-6.4pt} \slash}\;}
\def\Ds {{D \hspace{-6.4pt} \slash}\;}
\def\Gts {{\Gt \hspace{-6.4pt} \slash}\;}
\def\ds {{\del \hspace{-6.4pt} \slash}\;}
\def\ss {{\s \hspace{-6.4pt} \slash}\;}
\def\ks {{ k \hspace{-6.4pt} \slash}\;}
\def\ps {{p \hspace{-6.4pt} \slash}\;}
\def\xs {{x \hspace{-6.4pt} \slash}\;}
\def\pas {{{p_1} \hspace{-6.4pt} \slash}\;}
\def\pbs {{{p_2} \hspace{-6.4pt} \slash}\;}
\def\cFs {{{\cal F} \hspace{-6.4pt} \slash}\;}
\def\Dss {{D \hspace{-7.5pt} \slash}\;}
\def\dss {{\del \hspace{-7.0pt} \slash}\;}

\def\Ah{{\hat{A}}}
\def\Dh{{\hat{D}}}
\def\Gh{{\hat{G}}}
\def\Fh{{\hat{F}}}
\def\Ih{{\hat{I}}}
\def\Jh{{\hat{J}}}
\def\Kh{{\hat{K}}}
\def\Lh{{\hat{L}}}
\def\Ph{{\hat{P}}}
\def\Rh{{\hat{R}}}
\def\Vh{{\hat{V}}}
\def\Xh{{\hat{X}}}

\def\ah{{\hat{\a}}}
\def\bh{{\hat{\b}}}
\def\gh{{\hat{\g}}}
\def\dh{{\hat{\d}}}
\def\rh{{\hat{\r}}}
\def\hh{\hat{h}}
\def\uh{\hat{u}}
\def\xh{\hat{x}}
\def\yh{\hat{y}}
\def\ph{\hat{p}}
\def\xih{\hat{\xi}}
\def\chih{\hat{\chi}}
\def\Psih{\hat{\Psi}}
\def\phih{\hat{\phi}}

\def\psit{\tilde{\psi}}
\def\Psit{\tilde{\Psi}}
\def\Psibt{\tilde{\bar{Psi}}}

\def\lambdat{\tilde {\lambda}}
\def\st{\tilde{\sigma}}

\def\delt{\tilde{\delta}}
\def\Phit{\tilde{\Phi}}
\def\Phitb{\overline{\tilde{Phi}}}
\def\tht{\tilde{\th}}
\def\lt{\tilde{\l}}
\def\chit{\tilde{\chi}}
\def\phit{\tilde{\phi}}

\def\At{\tilde{A}}
\def\Bt{\tilde{B}}
\def\Ct{\tilde{C}}
\def\Dt{\tilde{D}}
\def\Et{\tilde{E}}
\def\Ft{\tilde{F}}
\def\Gt{\tilde{G}}
\def\Ht{\tilde{H}}
\def\It{\tilde{I}}
\def\Jt{\tilde{J}}
\def\Qt{\tilde{Q}}
\def\Rt{\tilde{R}}
\def\Mt{\tilde{M }}
\def\Nt{\tilde{N}}
\def\St{\tilde{S}}
\def\Vt{\tilde{V}}
\def\Xt{\tilde{X}}
\def\at{\tilde{a}}
\def\dt{\tilde{d}}
\def\htt{\tilde{h}}
\def\ft{\tilde{f}}
\def\gt{\tilde{g}}
\def\pt{\tilde{p}}
\def\qt{\tilde{q}}
\def\vt{\tilde{v}}
\def\nt{\tilde{n}}
\def\ut{\tilde{u}}
\def\wt{\tilde{w}}
\def\zt{\tilde{z}}
\def\xt{\tilde{x}}
\def\yt{\tilde{y}}
\def\Psit{\tilde{\Psi}}
\def\vphit{\tilde{\varphi}}
\def\tD{\tilde{\D}}

\def\eb{\bar{\epsilon}}
\def\delb{\bar{\partial}}
\def\thb{\bar{\theta}}
\def\mub{\bar{\mu}}
\def\lamb{\bar{\l}}
\def\psib{\bar{\psi}}
\def\sb{\bar{\sigma}}
\def\xib{\bar{\xi}}
\def\chib{\bar{\chi}}

\def\Psib{\bar{\Psi}}
\def\Phib{\bar{\Phi}}
\def\Lamb{\bar{\Lambda}}
\def\Sb{{\overline \Sigma}}
\def\cb{\bar{c}}
\def\hb{\bar{h}}
\def\qb{\bar{q}}
\def\wb{\bar{w}}
\def\ub{\bar{u}}
\def\zb{{\bar{z}}}
\def\Hb{\bar{H}}
\def\Qb{{\bar Q}}
\def\Omegab{\overline{\Omega}}
\def\ob{\overline{\omega}}

\def\Ab{{\overline A}} \def\Bb{{\overline B}} \def\Cb{{\overline C}}
\def\Db{{\overline D}} \def\Eb{{\overline E}} \def\Fb{{\overline F}}
\def\Gb{{\overline G}}
\def\Ib{{\overline I}}
\def\Jb{{\overline J}} \def\Kb{{\overline K}} \def\Lb{{\overline L}}
\def\Mb{{\overline M}} \def\Nb{{\overline N}} \def\Ob{{\overline O}}
\def\Pb{{\overline P}}  \def\Rb{{\overline R}}
 \def\Tb{{\overline T}} \def\Ub{{\overline U}}
\def\Vb{{\overline V}} \def\Wb{{\overline W}} \def\Xb{{\overline X}}
\def\Yb{{\overline Y}} \def\Zb{{\overline Z}}

\def\fb{{\overline f}}
\def\gb{{\overline g}}
\def\nb{{\overline n}}
\def\mb{{\overline m}}
\def\lb{{\overline l}}
\def\yb{{\overline y}}

\def\ldel{{\overleftarrow{\del}}}
\def\rdel{{\overrightarrow{\del}}}
\def\ldeldel{{\overleftarrow{\del^2}}}
\def\rdeldel{{\overrightarrow{\del^2}}}
\def\ldelb{{\overleftarrow{\bar{\del}}}}
\def\rdelb{{\overrightarrow{\bar{\del}}}}
\def\ba{{\bf a}}
\def\bk{{\bf k}}
\def\bl{{\bf l}}
\def\bp{{\bf p}}
\def\bq{{\bf q}}
\def\br{{\bf r}}
\def\bt{{\bf t}}
\def\bu{{\bf u}}
\def\bv{{\bf v}}
\def\bx{{\bf x}}
\def\by{{\bf y}}
\def\bA{{\bf A}}
\def\bR{{\bf R}}
\def\bV{{\bf V}}

\def\bz{{\boldsymbol{\zeta}}}

\def\bone{{\bf 1}}

\def\va{{\vec a}}
\def\vk{{\vec k}}
\def\vp{{\vec p}}
\def\vq{{\vec q}}
\def\vx{{\vec x}}
\def\vy{{\vec y}}
\def\vu{{\vec u}}
\def\vv{{\vec v}}
\def \vH{{\vec H}}
\def \vg{{\vec g}}

\def\vs{{\vec \sigma}}
\def\vtau{{\vec \tau}}

\newcommand{\ov}[1]{\overrightarrow{#1}}

\def\frA{\mathfrak{A}}
\def\frB{\mathfrak{B}}
\def\frC{\mathfrak{C}}
\def\frD{\mathfrak{D}}
\def\frE{\mathfrak{E}}
\def\frF{\mathfrak{F}}
\def\frG{\mathfrak{G}}
\def\frH{\mathfrak{H}}
\def\frM{\mathfrak{M}}
\def\frN{\mathfrak{N}}
\def\frR{\mathfrak{R}}
\def\frW{\mathfrak{W}}

\def\fra{\mathfrak{a}}
\def\frb{\mathfrak{b}}
\def\frf{\mathfrak{f}}
\def\frg{\mathfrak{g}}
\def\frh{\mathfrak{h}}
\def\frl{\mathfrak{l}}
\def\frs{\mathfrak{s}}
\def\fri{\mathfrak{i}}
\def\frj{\mathfrak{j}}

\def\ma{\mathfrak{a}}
\def\mg{\mathfrak{g}}
\def\mh{\mathfrak{h}}
\def\mR{\mathfrak{R}}
\def\mN{\mathfrak{N}}

\newcommand{\nn}{{\nonumber}}

\def\d{\delta}\def\D{\Delta}\def\ddt{\dot\delta}

\def\pa{\partial} \def\del{\partial}
\def\xx{\times}
\def\uno{\mbox{1 \kern-.59em {\rm l}}}

\def\trp{^{\top}}
\def\inv{^{-1}}
\def\dag{\dagger}
\def\pr{^{\prime}}

\def\rar{\rightarrow}
\def\lar{\leftarrow}
\def\lrar{\leftrightarrow}

\newcommand{\0}{\,\!}      %this is just NOTHING!
\def\one{1\!\!1\,\,}
\def\im{\imath}
\def\jm{\jmath}

\newcommand{\tr}{\mbox{tr}}
\newcommand{\slsh}[1]{/ \!\!\!\! #1}

\newcommand{\1}{\mbox{1}\hspace{-0.25em}\mbox{l}}

\def\vac{|0\rangle}
\def\lvac{\langle 0|}

\def\hlf{\frac{1}{2}}
\def\ove#1{\frac{1}{#1}}
\newcommand{\hot}[1]{\frac{#1}{2}}

\def\Box{\square}
\def\CC {\mathbb{C}}
\def\FF {\mathbb{F}}
\def\RR{\mathbb{R}}
\def\NN{\mathbb{N}}
\def\ZZ{\mathbb{Z}}
\def\bb#1{{\bf #1}}
\def\bcomment#1{}
\def\bfhat#1{{\bf \hat{#1}}}
\def\VEV#1{\left\langle #1\right\rangle}

\newcommand{\ex}[1]{{\rm e}^{#1}} \def\ii{{\rm i}}

\newcommand{\lrbrk}[1]{\left(#1\right)}
\newcommand{\lrsbrk}[1]{\left[#1\right]}
\newcommand{\sfrac}[2]{{\textstyle\frac{#1}{#2}}}

\def\stw{{\sqrt{2}}}

\def\rf {{\rm f}}
\def\ri {{\rm i}}
\def\rj {{\rm j}}
\def\rn {{\rm n}}
\def\rk {{\rm k}}
\def\rl {{\rm l}}
\def\rr {{\rm r}}
\def\rs {{\scriptscriptstyle \rm S}}
\def\rt {{\scriptscriptstyle \rm T}}

\def\rQ {{\scriptscriptstyle \rm \cQ}}
\def\rR {{\scriptscriptstyle \rm \cR}}

\def\cQb{{\cal \Qb}}
\def\cRb{{\cal \Rb}}
\def\cWb{{\cal \Wb}}

\def\fd {{\rm N}}
\def\afd {{\overline{\rm N}}}

\def \II {I\hspace{-.1em}I\hspace{.1em}}
\def \IIA {\mbox{\II A\hspace{.2em}}}
\def \IIB {\mbox{\II B\hspace{.2em}}}
\def \gs {g^s}
\def \ls {\lambda^s}

\def \I {{\cal I}}
\def \qs {q\hspace{-.53em}/\hspace{.15em}}
\def \ks {k\hspace{-.53em}/\hspace{.15em}}
\def \YM {{\mbox{\tiny YM}}}
\def \gym {g_{\YM}}

\def \Lc {\L_c}
\def\IR{\relax{\rm I\kern-.18em R}}
\def \id {{\bf 1}}

\def\cci{\ell}
\def\ccj{\ell'}

\def\bbq{\pmb{q}}
\def\bom{\pmb{\o}}
\def\bJ{\pmb{J}}
\def\bM{\pmb{M}}
\def\bB{\pmb{B}}
\def\bn{\pmb{n}}
\def\bE{\pmb{E}}

\newcommand{\rrr}[1]{\vskip 0.2cm \noindent{\bf #1} ---}

\long\def\symbolfootnote[#1]#2{\begingroup%
\def\thefootnote{\fnsymbol{footnote}}\footnote[#1]{#2}\endgroup}
\long\def\RemarkBox#1{\begin{flushleft}\fbox{\begin{minipage}
{17.5cm}{\bf Remark:} ~#1\end{minipage}}\end{flushleft}}
\newcommand{\aei}{\it Max Planck Institute for Gravitational Physics
(Albert Einstein Institute)\\ Am M\"uhlenberg 1, 14476 Golm,
Germany}

\newcommand{\nthu}{{\it Department of Physics, National Tsing-Hua
  University,
  Hsinchu 30013, Taiwan}}

\newcommand{\ctc}{{\it
Center of Theory and Computation, 
National Tsing-Hua University, Hsinchu 30013, Taiwan}}

\newcommand{\ictsustc}{{\it Interdisciplinary Center for Theoretical Study,
University of Science and Technology of China,\\
Hefei, Anhui 230026, People's Republic of China}}

\newcommand{\sysu}{{\it School of Physics and Astronomy, Sun Yat-Sen
    University, Zhuhai 519082, China}}

\newcommand{\ncts}{{\it
    National Center for Theoretical Sciences, Taipei 10617, Taiwan}}

\begin{document}

\begin{center}
~\vspace{20pt}
  
\thispagestyle{empty}
              {\Large \bf Tunneling, Page Curve and Black Hole Information}
\vspace{25pt}
 
Chong-Sun Chu ${}^{2,3,4}$\symbolfootnote[1]{Email:~\sf
  cschu@phys.nthu.edu.tw}, Rong-Xin Miao
${}^1$\symbolfootnote[2]{Email:~\sf  miaorx@mail.sysu.edu.cn}

\vspace{10pt}${{}^{1}}$\sysu \symbolfootnote[3]{All the Institutes of authors
  contribute equally to this work, the order of Institutes is adjusted
  for the assessment policy of SYSU.}

\vspace{5pt}${{}^{2}}$\ctc

\vspace{5pt}${{}^{3}}$\nthu

\vspace{5pt}${{}^{4}}$\ncts

\vspace{1cm}

\begin{abstract}
  In a recent paper \cite{Chu:2023mqi},  we proposed that the quantum states of
  black hole responsible for the
  Bekenstein-Hawking entropy are given by Bell states of Fermi quanta in the
  interior of black hole.
  In this paper, we
  include the effect of tunneling on these entangled states 
  and show that
  partial tunneling of these Bell states of Fermi quanta give rises to
  the Page curve
  of Hawking radiation. We also show that
  the entirety of information initially stored
  in the black hole is returned to the outside via the Hawking
  radiation.

\end{abstract}

\end{center}

\newpage
\setcounter{footnote}{0}
\setcounter{page}{1}
%%%%%%%%%%%%%%%%%%%%%%%%%%%%%%%

\tableofcontents

\setcounter{footnote}{0}

\section{Introduction}
There are mounting theoretical evidences that a black hole obeys the
first law of thermodynamics with an entropy
\be \label{BH}
S_{\rm BH} = \frac{A}{4G}
\ee
and
a temperature $T_H =1/(8 \pi G M)$ due to a thermal Hawking radiation
\cite{Hawking:1975vcx}.  Despite remarkable progress that has been
made in microstate counting \cite{Strominger:1996sh}, it is still not
known what is the nature of the gravitational degrees of freedom that
are being counted by \eq{BH}.

Another outstanding problem of the black hole is the information
problem \cite{Hawking:1976ra}.  Consider a black hole form from a pure
state. Assuming that the Hawking radiation is thermal, then the
entanglement entropy outside the black hole increases monotonically
for the entire course of life of the black hole. This {\it Hawking
  curve} violates the fundamental unitarity principle of quantum
mechanics: the fine-grained entanglement entropy should not exceed the
coarse-grained black hole entropy.  On the other hand, Page argued
that \cite{Page:1993wv,Page:2013dx} if the black hole evolution
process is unitary, then the total system of black hole and radiation
must go back to a pure state at the end of the evaporation.  As a
result, unitarity of quantum black hole requires the following
properties for the Hawking radiation:
1. Page curve: the entanglement entropy of Hawking radiation should
initially rise until the so called Page time $t_P$ when it starts to
drop down to zero as the black hole completely vanishes.
2. Recovery of information: at the end of the black hole life, one
must be able to recover from the Hawking radiation all the information
of the initial pure state.  It was Page's remarkable insight that the
Page curve behavior of the Hawking radiation, due to its accessibility
to external observers, can be especially useful as a decisive criteria
for unitary theory of quantum gravity.

Remarkable progress has been made lately on this front
where a fully quantum form \cite{Engelhardt:2014gca}
of the Ryu-Takayanagi entropy formula  \cite{Ryu:2006bv,Hubeny:2007xt}
was employed to compute the entanglement entropy
of the Hawking radiation and it was shown to obey the Page curve
\cite{Penington:2019npb,Almheiri:2019psf,Almheiri:2019hni}.
  Central to this analysis is the emergence of island,
  regions of spacetime that are completely disconnected and
spacelike separated from the region of the Hawking radiation. 
From  the semi-classical point of view,
island originated from the  wormhole saddle in the replica path
integral for the entanglement entropy \cite{Almheiri:2019qdq}.
It is interesting to note that the island is located
just beneath the horizon for evaporating Schwarzschild black hole
\cite{Penington:2019npb,Almheiri:2019psf,Almheiri:2019hni}.

That the Page curve is obtained gives confidence that the AdS/CFT
correspondence and the generalized entropy formula constitutes a
credible quantum gravity framework to study black holes. Yet, it still
leaves many questions unanswered.  For example, it is believed that
the Bekenstein-Hawking entropy \eq{BH} is coarse-grained in nature
\cite{Engelhardt:2017aux,Almheiri:2020cfm}. From this perspective,
what role does island play in coarse graining the initial pure state?
How does it help in returning the information to the environment?  The
main motivations of this work have been to understand better the
origin of the Page curve and the return of information in terms of
explicit spacetime quantum mechanical processes.

%v1 
In an earlier paper \cite{Chu:2022ieq}, it was proposed that the quantum states
of black hole responsible for the Bekenstein-Hawking entropy
are given by a thin shell of entangled quanta located at the region
just underneath the horizon. It was  argued
that the configuration can be stabilized by a new kind of
degeneracy pressure. However many questions were left unanswered. For example
what is the nature of these quanta? how does the
degeneracy pressure arises? 
In a recent paper \cite{Chu:2023mqi}, a model of the black hole
%c5 interior
spacetime as a system
of Fermi quanta was proposed. Due to the Pauli exclusion principle,
the system exhibits a quantum degeneracy pressure which is able to withstand
the collapse of spacetime due to gravity.
The horizon radius of the black hole is reproduced as a result of
the equivalence principle.
Moreover the Bekenstein-Hawking entropy is
explained in terms of the coarse graining of the
entangled configurations of the Fermi quanta.
According to the proposal,
the fermionic quanta observes a constant density of states, from which
it follows that there can be no more than $V/l_P^3$ quantum states in any
volume $V$. This maximal capacity of states implies the existence of a
minimal volume in space and suggests that any spacetime singularity as
predicted in general relativity must be resolved in the quantum gravity.

%v1
We note that these fermionic states in our model are characterized by energy
and are not localized in location as assumed in \cite{Chu:2022ieq}. Nevertheless
the analysis performed in \cite{Chu:2022ieq} can be easily adapted to the current consideration
to include the effect of  tunneling on the 
dynamical evolution of the black hole Fermi gas system.
In section 2, we
review our quantum mechanical model of black hole model as a gas of
Fermi quanta. In section 3, we
show that partial tunneling of the Bell states give rises to the Page curve
of Hawking radiation. We also 
show that at the end of life of the black hole,
the entirety of information initially stored in the black hole
is   returned to the outside via the Hawking radiation. Discussion and
conclusion are in section 4.

\section{Black hole as a system of Fermi quanta}

General relativity predicts that any object that has collapsed beyond
a certain point would form a black hole, inside which there is a
singularity. For a spherical black hole with horizon radius $R$, the
spacetime is given by some classical metric solution of the Einstein
equation.  And the singularity at the origin represents a place where
matter is compressed infinitely and the classical description of
spacetime breaks down.  In \cite{Chu:2023mqi}, a quantum mechanical model the
interior of the black hole is proposed. According to the model, the
spacetime inside the black hole is quantized whose fundamental quanta
are fermionic. The system of Fermi gas has a Fermi energy $\mu$ that
is inversely proportional to the horizon size $R$ and the density of
states is constant given by
\be
\mu = \frac{1}{2 \pi R}, \qquad g(E) = \frac{3\pi V}{G}. 
\ee
This gives the total number $N_S = \int dE g(E)$ of fermionic states  in
the system 
\be
N_S = \frac{2\pi R^2}{G},
\ee
which show an area dependence, 
and the total quantum energy of the Fermi sea is
\be
U = \frac{R}{2G}. 
\ee
When the black hole is charged or with a
cosmological constant, there will be additional contribution to the
internal energy of the Fermi system. It was shown in \cite{Chu:2023mqi} that
either by the equivalence principle, or equivalently, by the
requirement of the balancing of matter and gravitational pressure, the
black hole horizon radius is found to be reproduced remarkably.
Moreover, it was
shown that if there is a double bosonic degeneracy  for each energy level, then
the fermionic degrees of freedom at each energy level can
be entangled in 4 different ways, which then give rises to the
ground state degeneracy 
$\cG = 4^{N_L}$.
Here $N_L$ is the number of energy levels. It is related to the number of
states $N_S$ as $N_S = 4N_L$.
The counting of these microstates gives the entropy $S = \log \cG =  N_S/2$,
which reproduces precisely the Bekenstein-Hawking entropy.
A base 2 logarithm has been employed.
It is remarkable
that in our model, the holographic nature of the black hole 
% v2 is 
is a direct result of the area dependence of the
number of fermionic energy states of the black hole interior.

\section{Black hole information problem}

\subsection{Hawking radiation and unitarity violation}

Black hole information problem refers to the apparent violation of
unitarity of black hole in the semiclassical analysis of it's evolution.
There are at least two ways to see it,
both related to the Hawking radiation.
Consider a black hole 
% v2 form 
that forms
in flat space from a pure state and decays under Hawking radiation.
If unitarity of
quantum mechanics is not violated in the black hole formation process,
then the black hole interior
and  exterior together is in a pure state. 
Due to our ignorance of the interior, this give rises to a number
$S_{\rm BH}$ of coarse-grained states with entropy \eq{BH}.
By definition, the knowledge of these states together with the knowledge
of the exterior constitutes a pure state. Eventually,
the black hole is exhausted by the Hawking radiation. According to
Hawking's original calculation, Hawking radiation is thermal and 
a mixed state. 
% v2 This is however
This is, however, in direct contradiction with unitarity
since a pure state cannot evolve into a mixed state.
Instead if unitarity is preserved, then the final state of the Hawking
radiation cannot be purely thermal, but it must encompass the information
contained in the initial set of coarse-grained states so that,
together with the outside information, a pure state
can be reconstructed.

A related problem is the behavior of the entropy
of the Hawking radiation. An easy way to see this is to consider
the quantum mechanical picture of  Parikh and Wilczek
\cite{Parikh:1999mf},
where 
Hawking radiation is obtained from the tunneling
of the virtual particles created out of the vacuum state of the black hole.
For an outgoing particle with energy $\o$ created just inside the
horizon, they found that it can travels cross the horizon and results
in a nonvanishing imaginary part for the particle action: $ {\rm Im} S
= 4 \pi \o (M-\frac{\o}{2}).  $ This corresponds to a tunneling
process with the tunneling rate
\be \label{tr}
\G = \G_0 e^{-2 {\rm Im} S} = \G_0
e^{-8\pi \o (M-\frac{\o}{2})},
\ee
where $\G_0$ is a prefactor which can
be computed from a more detailed knowledge of the dynamics.  The
leading exponential $\o$ dependence in $\G$ registers a Boltzmann
thermal distribution with the Hawking temperature $T$. The higher order correction
$\o^2$ term is a back reaction term, showing that the spectrum is slightly
deviated from the thermal one.
%v1
We note that such a result for the tunneling rate is more or less
guaranteed by consistency due to the fact that Hawking temperature is
an universal quantum effect of the horizon, which can be understood either in
terms of the removal of conical singularity of geometry or as Unruh
effect of acceleration.   In addition, pair creation outside the
horizon was also considered.  In this so called anti-particle channel,
the anti-particle
%v1
created outside the horizon
follows a time reversed ingoing geodesic crosses the
horizon and also makes contribution to the Hawking radiation.
Note  that while
the original analysis of \cite{Parikh:1999mf} is for massless
particles, one can show that the same result \eq{tr} holds for massive
particles as well.

%v1 
Although this picture of tunneling gives a clear physical picture of the
origin of Hawking radiation and its temperature, it also leads to
the Hawking curve of the entropy of Hawking radiation.
To see this, consider a virtual pair created in
the interior side of the horizon (particle channel).  Since the pair
was created entangled, an entanglement between the black hole and the
Hawking radiation is created when the particle tunnels through and the
anti-particle got absorbed by the black hole.  The rate of increase of
the entanglement entropy of the Hawking radiation is given by $\a (\o)
A$, where $\o$ is the energy of the particle and $\a(\o)$ is the
creation rate per unit area of virtual pairs on the interior side of
horizon times the tunneling probability.  Similarly, there is a
contribution $\b (\o) A$ from the anti-particle channel, where
$\b(\o)$ is the creation rate of virtual pairs on the exterior side of
horizon times the tunneling probability.  As $\a, \b$ are positive,
the entanglement entropy of Hawking radiation increases
monotonically. This Hawking curve violates the unitarity of quantum
mechanics and is a fundamental problem for the quantum information of
black hole.

%v1
It has been suggested that some form of nonlocality is
needed to resolve the black hole
information problem (see \cite{Giddings:2006sj} for a review).  
As quantum nonlocality is best captured by entanglement, it seems natural
to consider the coarse-grained degrees of freedom to be rich in entanglement
content. This is exactly what happens in our black hole model.
In this section, we include the dynamical
effect of  tunneling and 
show how tunneling can affect Fermi gas of quanta
and lead to the Page curve and the recovery of the full entanglement
content of the coarse-grained states in late time correlation of the
Hawking radiation.

\subsection{Tunneling and entanglement swapping of Bell pairs in the Fermi
  model}

%v1 
Our model of black hole differs from the quantum vacuum model of black hole
in a major way.
Instead of a vacuum filled with
virtual pairs of particle anti-particle \cite{Parikh:1999mf},
the black hole in our picture is
filled with a collection of entangled fermionic energy eigenstates.
In the  picture of \cite{Parikh:1999mf}, it is
the motion of particles over the classically forbidden trajectories that
give rises to the tunneling and the Hawking radiation.
Our fermionic states are not localized in space and the mechanism for their
tunneling has to be different. Nevetheless, we do expect that
in quantum gravity, the Fermi quanta will also 
be able to tunnel and leak through the horizon.
Although we do not currently have the
technology of quantum
gravity to determine the tunneling rate, we can expect the tunneling rate
will take the generic form \eq{tr}
\footnote{A quantum mechanical
model of quantum gravity  was recently proposed in \cite{Chu:2024qil}
wherein a quantum black hole
is represented by a non-commutative geometry and is populated by a Fermi sea.
These solutions are classically stable but can decay by tunneling. It is
interesting to study in detail the tunneling process in the quantum mechanical
model.}
In any case, our following analysis and conclusion 
will be generic and will not depend
on the detailed form  of the tunneling rate.

In addition to tunneling of the existing Fermi quanta, there will also be
vacuum pair creation process as usual and this leads to an  interesting
effect of {\it entanglement swapping} on the Fermi quanta.
Consider the particle channel, the anti-fermionic quanta $\bar{p}$
that is left behind can get annihilated by a quanta $b_2$
of one of the Bell pairs.
As a result, the fermionic $p$
becomes entangled with the other
partner $b_1$ of the Bell pair, and entanglement is swapped from
the Bell pair and the virtual pair to one between the
Hawking radiation quanta and the
remaining quanta inside the black hole.
Similarly, 
entanglement swapping occurs in the anti-particle channel.
See  Figure 1.
Both these processes will not only alter the entropy content of the black
hole, but also introduce an  entanglement entropy for the Hawking
radiation due to the partially escaped  entangled pairs.
\begin{figure}
  \label{bell}
  \centering
\includegraphics[width=0.3\linewidth]{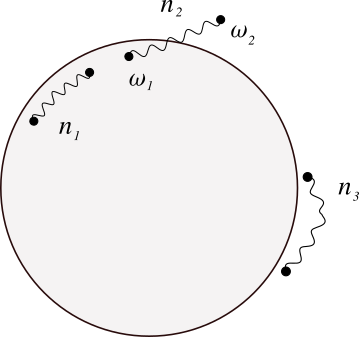}
\caption{Tunneling of
entangled quanta}
\end{figure}

To study the dynamics of the
entangled states,
let us
denote the energies of the quanta of each
entangled pair by
%v1 
$0\leq \o \leq M_0$, where $M_0$ is the original mass of the black hole.
Let
$n_1 (\o,t) d\o$ be  the number of  entangled pairs inside the black hole
and have energies within the interval $(\o, \o+d\o)$,
$n_2(\o, t) d\o $
be the number of entangled pairs that has
one of the quanta
inside the horizon and the other outside, and
 $n_3(\o, t) d\o $ be
the number of
entangled pairs  that are located entirely outside the horizon.
%v1 In order to avoid over counting,
%% the domain of energies for $n_1(\o_1,\o_2), n_3(\o_1,\o_3)$  is given by
%% $D := \{ (\o_1,\o_2) |\;  0\leq \o_1 \leq \o_2 \leq M_0 \}$.
%% As for $n_2(\o_1,\o_2,t)$, the
%% first (resp. second) argument $\o_1$ (resp. $\o_2$)
%% refers to the energy of the particle that
%% is inside (resp. outside)
%% the horizon. There is no constraint on the size of
%% $\o_1, \o_2$ and the
%% domain for $n_2$ is given by
%% $ D' := \{ (\o_1,\o_2) | \;  0\leq \o_1, \o_2 \leq M_0 \}$.
See Figure 2. 
Taking into account of tunneling
of the Bell states and quantum entanglement swapping,
%v1 it is easy to obtain
we have
\begin{align}
  \frac{\del n_1}{\del t} &= - 2 \G \, n_1 - 2\d A,
  \label{n1}\\
  \frac{\del n_2}{\del t} &=  2\G n_1 -\G n_2 + 2 \d A,
  \label{n2} \\
%  \quad \mbox{for $\o_1\leq \o_2$}, \label{n2a}\\
% \frac{\del n_2}{\del t} &=  \G_2 \nt_1 -\G_1 n_2 + \d_1 A,
%\quad \mbox{for $\o_1 > \o_2$}, \label{n2b}
  \frac{\del n_3}{\del t} &=  \G n_2 ,
  \label{n3}
\end{align}
where
$\G := \Gamma(\o,M(t))$ 
is the tunneling rate for a Fermi quanta of
energy $\o$ from a
black hole of mass $M(t)$.
$\a :=\a(\o,M(t))$ (resp. $\b :=\b(\o,M(t))$)
is the production rate per unit area of the conventional
vacuum created Hawking radiation in the anti-particle channel (resp. particle
channel), and $\d:=\a+\b$ is the total
production rate from both channels.
%v1 Here $n_i = n_i(\o_1,\o_2,t)$,
%% while $\nt_{1,2} := n_{1,2} (\o_2,\o_1,t)$ has its arguments reversed. 
Physically, the $n_i$ terms and the $A$ terms on the RHS of \eq{n1}
(resp. \eq{n2} or \eq{n3}) represent the effect of direct tunneling
and quantum swapping on the entangled pairs that are entirely inside
the black hole (resp.  partially inside or entirely outside the black
hole). The factors of 2 in \eq{n1}, \eq{n2} is due to the fact that both ends
of the entangled pair can participate in these processes. 
The total number of each types of
entangled pairs is
\be
N_i (t) = \int_D d\o  n_i (\o, t),\quad i=1,2,3,
\ee
where
the integration region is over  $D := \{0 \leq \o \leq M (t)\}$.
Physically, $2N_1$ 
represents
the amount of coarse-grained entropy of the black hole at time $t$,
$N_2$ represents the entanglement entropy of Hawking radiation, and
$N_3$ represents the amount of entanglement information contained in the
Hawking radiation.
In our model, the mass of the black hole is given by
\be \label{Mt}
M(t) = \int_D d\o \left(
  2 n_1 + n_2
\right) \o
  \ee
  where the appropriate domains of integration is understood. In a
  consistent analysis, \eq{Mt} will have to be considered together
  with \eq{n1} - \eq{n3}. This is quite a complicated system to solve.
  In  the following, we show that without assuming the form of $\G(\o,M)$
  and  $\d(\o,M)$,
  and without solving the system explicitly, one can immediately show
  that the Hawking radiation obeys the Page curve and that the
  complete information of the black hole is returned.
\begin{figure}
   \label{swap}
   \centering
\psfrag{P}{$p$}
\psfrag{AP}{$\bar{p}$}
\psfrag{b1}{$b_1$}
\psfrag{b2}{$b_2$}
\hspace{-1cm}
\includegraphics[width=0.6\linewidth]{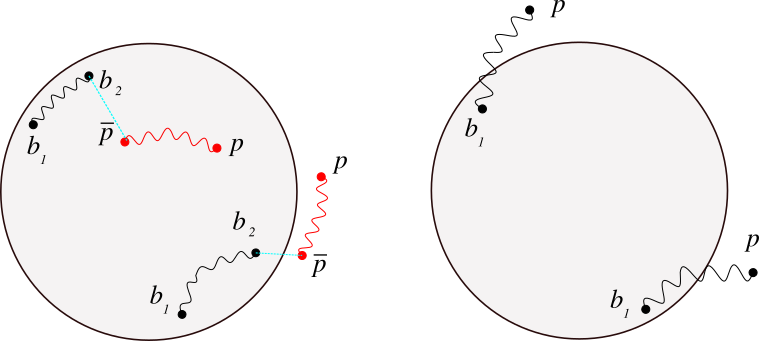}
\caption{Entanglement swapping induced by virtual pairs}
\end{figure}

Let us consider a black hole formed at $t=0$, implying
the conditions $n_2 =n_3 =0$ initially.
Consider first \eq{n1}. As the RHS of \eq{n1} is 
non-positive, 
$n_1$ will continue to decrease until
it reaches zero, where also $\del n_1/\del t =0$.
Denote this time as $t_{1,\o}$ and define
$t_1 : = {\rm Supp}_{\o\in D} (t_{1,\o})$.
We have
\be
N_1=0 \;\;\; \mbox{for $t\geq t_1$}.
\ee
Next  consider \eq{n2}. Starting with the initial conditions
$n_2=0$ and $n_1 >0$, $n_2$ is set to increase initially.
As time progress, there will be a cross over time  where the RHS of
\eq{n2} vanishes and turns negative subsequently. Then
$n_2$ will continue to decrease until it reaches zero
at some time $t_{2,\o}$.
Define
$ t_E : = {\rm Supp}_{\o\in D} (t_{2,\o})$,
we have
\be
N_2 =0 \;\;\; \mbox{for $t\geq t_E$}.
\ee
Finally  consider \eq{n3}. It is clear that $n_3$ increases monotonically
until $n_2$ reaches zero at $t=t_{2,\o}$, then it remains a constant.
Therefore we have
\be
\mbox{$N_3 = N_{3f}$ a constant}\;\;\; \mbox{for $t \geq t_E$}.
\ee
It is clear that $t_1$ is when the black hole's
degrees
of freedom become completely entangled with the environment
and
$t_E$ is the end time that the black hole dies.

Note that as
$N_2$ starts zero at $t=0$
and ends at zero again at $t=t_E$, it must
follow an inverted V-shape curve
and reaches a maximum at some
intermediate time $t_P$. This is generic and we obtain 
the Page
curve for the Hawking radiation, with $t_P$
being the Page time. 
By expressing
$N_2$ as integrals over the domain $D$, it is easy to establish
the conservation equation:
\be \label{cons}
\frac{d}{dt} (N_1+ N_2 + N_3) =0.
\ee
It is remarkable that this leads immediately to
\be
N_{3f} = N_{10},
\ee
meaning that all the entanglement
information originally stored in the
Bell pairs are
returned to the exterior observer via the Hawking radiation.
% v3 
%We note that it is crucial in
%our model to include the thin shell of Bell particles
%to start with and allow the tunneling of them.
%Without these,
Without the  tunneling of Bell particles,
our equations \eq{n1}-\eq{n3} will reduce to the equation
$\del n_2/\del t = \d_1 A$
as in the conventional tunneling model of Hawking radiation.
There would then be the Hawking curve and it
would not possible to return information
via $n_3$. 
In the supplementary material,
we
%v1
provide justification that $t_E$ is finite in our model.
We also show the explicit time dependence of the
$n_i$'s for some typical values of the energies, which confirm the generic
behavior
discussed here.
We also note that in our model, the Hawking radiation
is
%v1 non-thermal
not purely thermal
and contains correlations that
are
necessarily highly nonlocal as
they could
arise
from tunneling process that occurs at very different times
over the life of the black hole.

%v1  We remark that an infalling observer 
%% will encounter the thin shell of entangled particles
%% located underneath the horizon
%% and get assimilated there as new Bell particles. 
%% In this sense the
%% shell of
%% Bell particles acts
%% like a firewall \cite{Almheiri:2012rt}. 
%% We have considered and
%% proposed that the
%% wave function of the
%% thin shell matter can
%% be written in terms
%% of 2-qubit Bell states. It is interesting to understand
%% if this is true and
%% how the firewall makes it.

{\it Acknowledgments:} We thank Pei-Ming Ho for useful discussion.
C.S.C acknowledge support of this work by NCTS and
the grant 110-2112-M-007-015-MY3 of the National 
Science and Technology Council of Taiwan. 
R.X.M thank support by NSFC grant (No. 11905297).

\appendix

\section{Appendix}

In a semiclassical estimation,
Hawking radiation reduces the mass of the black hole as
\cite{Page:1976df}
\be \label{M-rate}
\frac{d M}{dt} = - \frac{Q}{3 M^2},
\ee
where $Q=3 \a/G^2$ and $\a$ is some
numerical constant. This gives
\be\label{M3}
M(t) = M_0 \left[ 1- \frac{Q t}{ M_0^3} 
\right]^{1/3}
\ee
and a finite
lifetime of the black hole
$ t_E = M_0^3/Q$.
In our model, this should come out from the consistent
set of equations
\eq{n1}-\eq{n3} and \eq{Mt}.
Although the complete analysis is quite
complicated, we can see that our model does roughly give
\eq{M-rate} near the final stage of evaporation and so $t_E$
is finite.
In fact, using \eq{n1} and \eq{n2}, we have
  \be\label{Mt-rate}
  \frac{d M}{dt} = \int_{D} - \G (2n_1+n_2) \o,
  \ee
  where we have ignored the much smaller term $- 2 \d \o A$
  on the RHS 
  since $\d \ll \G$ as $\d$ involves an additional vacuum creation rate.  
Near the final stage of evaporation, it is
$\int n_1  \sim O(1)$ and we can use the mean value theorem of
calculus to estimate that
$\int  \G \o n = \overline{ \; \G \o}  \, \int n
\sim 1/M^2$,
where we have used $\o \sim M$ and that $\G \sim \G_0$ for small $M$. Here
the prefactor of the tunneling
rate has a dependence $\G_0 \sim 1/M^3$ coming from the phase space volume. As
a result, \eq{Mt-rate} is consistent with the semiclassical result \eq{M-rate}
at time close to the end point
and so $t_E$ is finite.

To get a better feeling of the time evolution  of the Bell pairs,
let us consider the
approximate mass function $M(t)$ \eq{M3} with $M_0=1,Q=0.05$ so
that $t_E= 20$. In Figure 3,
\begin{figure}[ht]
  \centering
\includegraphics[width=8.5cm]{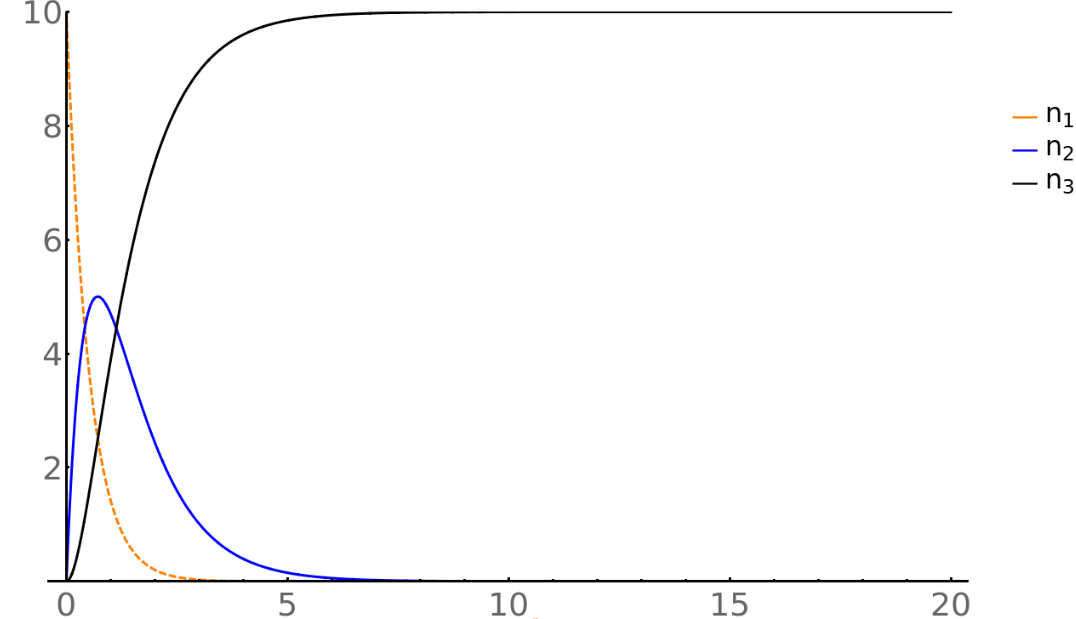}\\
\includegraphics[width=8.5cm]{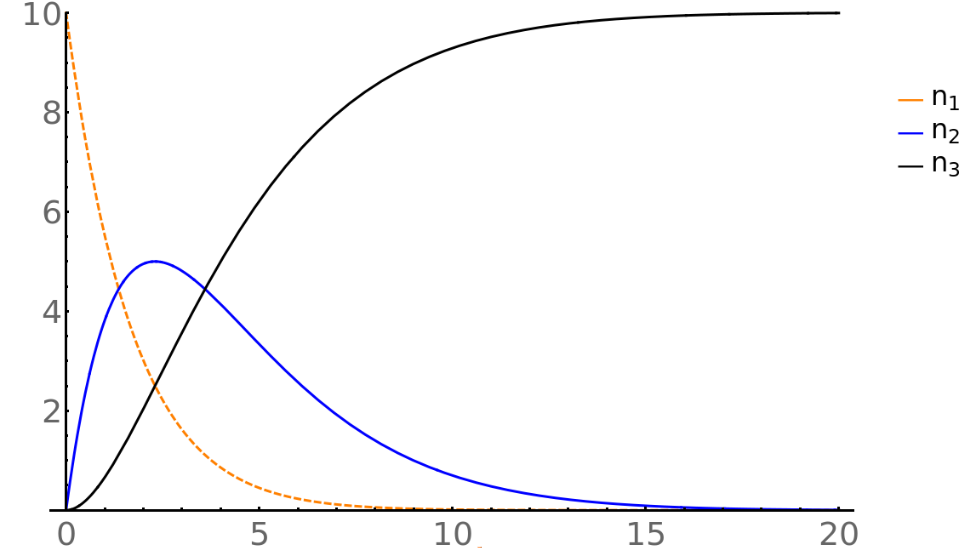}
\caption{Evolution of Bell states for the
  energies $\o = $
  (a) 0.001,
  (b) 0.05,
      The mass \eq{M3} with $M_0=1,Q=0.05$ is adopted for these plots.  }
\end{figure}
we show the plots of $n_1, n_2,  n_3$
for two typical set of energies  $\o$ given by
(a): 0.001 and
(b): 0.05.
Since $\d \ll \G$, we have taken $\d =0$  here for simplicity.
Including $\d\neq 0$
won't affect the qualitative features of these plots.
%v1 much.
We see that the generic behaviors of the $n_i$'s
discussed above do get captured
very well here, showing that \eq{M3} is a
decent approximation.
%v1 However, we can see that for the set (a) of energies,
% the time $t_2$ for $n_2$ and $\nt_2$ to decrease to zero is actually slightly
% larger than $t_E=20$, showing that \eq{M3} is not entirely consistent with
% \eq{n1}-\eq{n3}.
% Nevertheless \eq{M3} is a pretty good approximation and
%v1
In general, a perturbative
scheme can
%v4 in principle
be devised to solve the system \eq{n1}-\eq{n3}, \eq{Mt}
numerically.

\bibliographystyle{utphys}

\bibliography{references}

\end{document}